\documentclass[apj]{emulateapj}
\usepackage{amsmath}
\usepackage{makeidx, xcolor}
\definecolor{linkc}{rgb}{0,0,1.0}
\usepackage[breaklinks, colorlinks, urlcolor={linkc}, citecolor={linkc}, linkcolor={linkc}]{hyperref}
\usepackage{breakurl}
\usepackage{etoolbox}

\makeatletter

\patchcmd{\NAT@citex}
  {\@citea\NAT@hyper@{%
     \NAT@nmfmt{\NAT@nm}%
     \hyper@natlinkbreak{\NAT@aysep\NAT@spacechar}{\@citeb\@extra@b@citeb}%
     \NAT@date}}
  {\@citea\NAT@nmfmt{\NAT@nm}%
   \NAT@aysep\NAT@spacechar\NAT@hyper@{\NAT@date}}{}{}

\patchcmd{\NAT@citex}
  {\@citea\NAT@hyper@{%
     \NAT@nmfmt{\NAT@nm}%
     \hyper@natlinkbreak{\NAT@spacechar\NAT@@open\if*#1*\else#1\NAT@spacechar\fi}%
       {\@citeb\@extra@b@citeb}%
     \NAT@date}}
  {\@citea\NAT@nmfmt{\NAT@nm}%
   \NAT@spacechar\NAT@@open\if*#1*\else#1\NAT@spacechar\fi\NAT@hyper@{\NAT@date}}
  {}{}

\makeatother

\newcommand{\psr}{{B1509}}
\newcommand{\chandra}{{\it Chandra}}

\newcommand{\exosat}{{\it EXOSAT}}
\newcommand{\ginga}{{\it Ginga}}

\newcommand{\xmm}{{\it XMM-Newton}}

\newcommand{\rxte}{{\it RXTE}}

\newcommand{\fermi}{{\it Fermi}}

\newcommand{\nustar}{{\it NuSTAR}}
\newcommand{\bepposax}{{\it BeppoSAX}}

\begin{document}
\title{\emph{Chandra} Phase-Resolved Spectroscopy of the High Magnetic Field Pulsar B1509$-$58}


\author{Chin-Ping Hu$^1$, C.-Y. Ng$^1$, J. Takata$^2$, R. M. Shannon$^{3,4}$, S. Johnston$^4$}
\affiliation{$^1$ Department of Physics, The University of Hong Kong, Pokfulam Road, Hong Kong; {\color{blue} cphu@hku.hk, ncy@bohr.physics.hku.hk}\\
$^2$ School of Physics, Huazhong University of Science and Technology, Wuhan, Hubei, China \\
$^3$ International Centre for Radio Astronomy Research, Curtin University, Bentley, WA 6102, Australia \\
$^4$ CSIRO Astronomy and Space Science, Australia Telescope National Facility, Box 76, Epping, NSW 1710, Australia \\
}
\submitted{ApJ, in press} 

\shortauthors{Hu et al.}
\shorttitle{\emph{Chandra} Spectroscopy of PSR\,B1509$-$58}

\begin{abstract}
We report on a timing and spectral analysis of the young, high magnetic field rotation-powered pulsar (RPP) B1509$-$58 using \chandra\ continuous-clocking mode observation. The pulsar's X-ray light curve can be fit by the two Gaussian components and the pulsed fraction shows moderate energy dependence over the \chandra\ band. The pulsed X-ray spectrum is well described by a power law with a photon index 1.16(4), which is harder than the values measured with \rxte/PCA and \nustar. This result supports the log-parabolic model for the broadband X-ray spectrum. With the unprecedented angular resolution of \chandra, we clearly identified off-pulse X-ray emission from the pulsar, and its spectrum is best fit by a power law plus blackbody model. The latter component has a temperature of $\sim$0.14\,keV with a bolometric luminosity comparable to the luminosities of other young and high magnetic field RPPs, and it lies between the temperature of magnetars and typical RPPs. In addition, we found that the nonthermal X-ray emission of PSR B1509$-$58 is significantly softer in the off-pulse phase than in the pulsed phase, with the photon index varying between 1.0 and 1.8 and anticorrelated with the flux. This is similar to the behavior of three other young pulsars. We interpreted it as different contributions of pair-creation processes at different altitudes from  the neutron star surface according to the outer-gap model.
\end{abstract}


\section{Introduction}
Pulsars are fast-rotating neutron stars. Most of them spin down steadily, converting their rotational energy into electromagnetic radiation and particle outflows. This class of pulsars is thus known as rotation-powered pulsars (RPPs). They typically have spin periods of $\sim$0.1--10\,s and period derivatives of $\sim10^{-12}$--$10^{-17}$\,s\,s$^{-1}$, implying magnetic fields of $B\sim10^{11}$--$10^{13}$\,G. Thanks to the revolution of astronomical instrumentation in the past two decades, several subtypes of pulsars have been discovered, and they occupy different regions in the $P$-$\dot{P}$ diagram. Magnetars are an extreme group of pulsars with long spin periods of 2--12\,s and strong magnetic fields of $10^{14}$--$10^{15}$\,G inferred from the spin-down rate. They are remarkable for their energetic burst activities and high X-ray luminosities \citep[see review by][]{Mereghetti2008, MereghettiPM2015}. The theoretical interpretation is that the energetic features are powered by the dissipation and decay of a strong magnetic field \citep{ThompsonD1995, ThompsonD1996}, although the mechanism that converts the magnetic energy into the surface heating remains an open question \citep[see][]{TurollaZW2015, BeloborodovL2016}.  Moreover, recent discoveries blurred the boundary between magnetars and RPPs. The low field magnetars SGR 0418+5729 and Swift J1822.3$-$1606 have spin-down-inferred dipole magnetic fields of $6\times10^{12}$\,G and $1.4\times10^{13}$\,G, respectively, comparable to typical RPPs \citep{ReaET2010, ReaIP2013, ScholzNL2012}. Furthermore, the discovery of magnetar-like bursts of high magnetic field RPPs J1846$-$0258 with $B=4.9\times10^{13}$\,G \citep{GavriilGG2008, NgSG2008}, and J1119$-$6127 with $B=4.1\times10^{13}$\,G \citep{ArchibaldKT2016, GogusLK2016} also challenges the division between these two classes of neutron stars. To interpret the fact that magnetar and high magnetic field RPPs share similar behaviors, a unified model has been proposed \citep{PernaP2011, ViganoRP2013} based on the magneto-thermal evolution \citep[e.g.,][]{PonsG2007}. 

Because pulsars are born as hot objects, young pulsars provide good targets for measuring the thermal emission and testing the above theory. There are only three high magnetic field ($B>10^{13}$\,G) RPPs younger than 2000 years: PSRs\,B1509$-$58, J1846$-$0258, and J1119$-$6127. \object{PSR\,B1509$-$58} (hereafter B1509) is the only one for which the thermal emission has not been studied. To complete the sample, we carry out a measurement of the surface temperature of \psr\ with the \emph{Chandra X-ray Observatory} in this study.

\begin{figure}[ht]
\epsscale{1.15}
\plotone{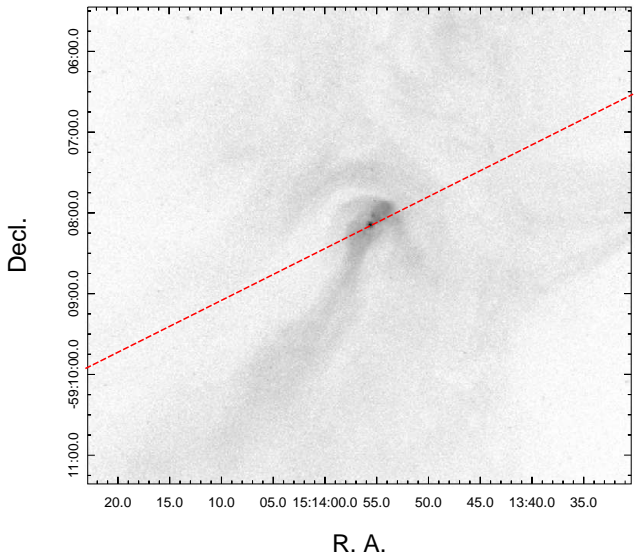} \caption{\chandra\ X-ray image of \psr\ and its PWN in MSH 15-5{\it 2} in the 0.5--7\,keV band. All the X-ray events are projected onto the red dashed line in the CC mode observation to form a 1D image. This image is created by accumulating all the archival observations with \chandra\ ACIS-I and operated in the TE mode. \label{chandra_image}}
\vspace{-0.4cm}
\end{figure}

\psr\ has a spin period of 150\,ms, which was first discovered in X-rays with the {\it Einstein Observatory} \citep{SeawardH1982}. It is embedded in a bright supernova remnant MSH 15$-$5{\it 2} \citep{MillsSH1960}. Radio pulsations were detected subsequently, and a large period derivative of $\sim 1.5\times 10^{-12}$ s\,s$^{-1}$ was also confirmed \citep{ManchesterTD1982, WeisskopfED1983}. This implies a dipole field of $B=1.5\times10^{13}$\,G, which is higher than that of SGR 0418+5729. The radio pulse profile shows a sharp peak, while the X-ray profile has a single, broad, and asymmetric peak and lags the radio peak by about one quarter of the cycle \citep{KawaiOB1991}.  The broad X-ray pulse profile was conventionally described by two Gaussian components \citep{CusumanoMM2001, GeLQ2012}. Using 11 years of radio timing observations, the detailed spin-down evolution suggests a characteristic age of 1700 years \citep{KaspiMS1994}. The distance is $5.2\pm1.4$\,kpc estimated from the H{\sc i} measurement \citep{GaenslerBM1999}.

\psr\ is associated with a bright pulsar wind nebula (PWN). The high-resolution \chandra\ image in Figure~\ref{chandra_image} shows a bright jet in the southeast and a semicircular arc 30\arcsec\ north of the pulsar \citep{GaenslerAK2002}. A few bright clumps were found in the northwest and are occasionally seen in the southeast, indicating the turbulent nature of the flows near the pulsar \citep{DeLaneyGA2006}. Moreover, an inner ring-like structure with a 10\arcsec\ radius surrounding the pulsar was found, which may correspond to the wind termination shock \citep{YatsuKS2009}. Although \chandra\ can resolve \psr\ from the  surrounding PWN, its surface temperature has never been determined because the emission is too bright, such that the spectrum was heavily distorted by the pile-up effect. \citep{GaenslerAK2002}.

Using the \exosat, \ginga, and SIGMA data, \citet{GreiveldingerCM1995} determined a hydrogen column density of the PWN of $N_{\rm H}\approx9.5\times10^{21}$\,cm$^{-2}$ and concluded that the pulsar had a harder emission than its surrounding nebula. The pulsed spectrum obtained with the \emph{RXTE} can be well fit by a power law (PL) with a photon index $\Gamma=1.36\pm0.01$.  The off-pulse spectrum, which was dominated by the surrounding nebular emission due to the poor spatial resolution, shows a much softer PL with $\Gamma=2.215\pm0.005$ \citep{MarsdenBG1997}.  A pulse phase-resolved spectral analysis showed that $\Gamma$ remains stable during the pulsed phase, such that the spectra of the two components of the pulse shape are indistinguishable \citep{RotsJM1998}. \citet{GeLQ2012} presented a more comprehensive phase-resolved spectral analysis using 15 years of \rxte\ data with a total exposure time of $\sim$578\,ks. They treated the 0.2 cycles off-pulse emission as the background and found that $\Gamma$ increases from 1.33 to 1.47 as the flux decreases. In addition, the photon indices observed with LECS, MECS, and PDS onboard \bepposax\ are significantly different and the 0.1--300\,keV broadband spectrum is better fit by a curved log-parabolic model than by a single PL \citep{CusumanoMM2001}. The recent \nustar\ observation provides supporting evidence for the log-parabolic model \citep{ChenAK2015}, although the parameters slightly deviate from those determined from \bepposax. However, both the PL and the log-parabolic models are equally good in a detailed phase-resolved analysis. 

In this paper, we present a detailed phase-resolved analysis of the X-ray emission of \psr, using high temporal resolution data taken with \chandra.  We describe the observations in Section \ref{observation}, including a new observation made with the ACIS-S in the continuous-clocking (CC) mode and an archival observation made with the High Resolution Camera (HRC).  Section \ref{analysis} presents the data analysis and results, including the detection of the off-pulse emission from the neutron star, the on- and off-pulse spectra, and the phase-resolved analysis.  We then discuss the results in Section \ref{discussion} and give a summary in Section~\ref{summary}.

\section{Observations and Data Reduction}\label{observation}
A new observation of \psr\ was made with \chandra\ on 2013 March 29 with a total exposure of 60.1\,ks \dataset[ ADS/Sa.CXO#obs/14805]{ObsID 14805}.  The ACIS-S CCD array was operated in the CC mode with a high timing resolution of 2.85\,ms, such that the pile-up is negligible.  Note that the CC mode observation only has one-dimensional spatial information since the events are collapsed along CCD columns into a line. The roll angle of the telescope was carefully chosen to avoid the bright features of the PWN, including the jet and the clumps, from falling onto the same CCD column as the pulsar (see Figure~\ref{chandra_image}).  We also used the archival observation \dataset[ ADS/Sa.CXO#obs/5515]{ObsID 5515} made with the HRC-I on 2015 June 13 to verify the detection of the off-pulse emission.

 
We reprocessed both the ACIS-S and HRC data using the task {\tt chandra\_repro} in the Chandra Interactive Analysis of Observations (CIAO) v4.8 with the most recent calibration database (CALDB) 4.7.0.  To perform an accurate timing analysis, we corrected all photon arrival times to the barycenter of the solar system with the CIAO tool {\tt axbary}, based on the JPL solar system ephemeris DE405. We examined the background light curve and found no significant background flares in either observations. Therefore, all the time spans with exposures of 59.9 ks for ACIS-S CC mode data and 44.9 ks for HRC data were used in this analysis. 

\begin{figure}[ht]
\centering
\includegraphics[bb=0 0 290 270, width=0.42\textwidth, clip]{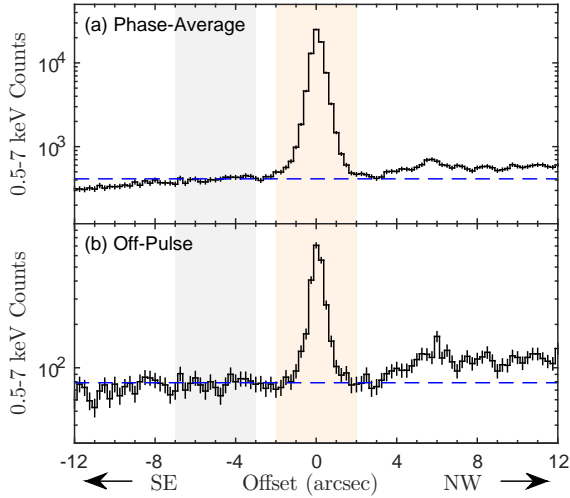} 
\caption{Projected photon distribution of \psr\ obtained with the \chandra\ CC mode observation in the 0.5--7\,keV energy range (see Figure~\ref{chandra_image}).  Data were collected from (a) the phase-averaged and (b) off-pulse phases. Positive offset is defined as the northwest direction from the pulsar. The orange shaded area indicates the source region and the gray shaded area indicates the background region.  The blue dashed line is the mean background value. \label{projected_counts}}
\end{figure}

\section{Data Analysis and Results}\label{analysis}
\setcounter{footnote}{0}
\subsection{Radio Ephemeris}
We used the pulsar ephemeris obtained with the Parkes Observatory, which is a part of the timing program for \fermi\ \citep{WeltevredeJM2010}. A total of 88 times of pulse arrivals (TOAs) between MJD 54220 and MJD 56512 were used to generate the short time-baseline ephemeris.  We then used the TEMPO2 package to fit the TOAs with the rotational frequency and its first two time-derivatives.  Because of the significant timing noise of \psr, we furthermore used the FITWAVES algorithm \citep{HobbsLK2004}, which employs the sinusoidal curves instead of higher-order polynomials, to fit the residuals.  We used five harmonics to whiten the timing noise, and the final rms residual is 1.4\,ms.  The dispersion measure was set to $252.5\pm0.3$\,pc\,cm$^{-3}$.  

\subsection{Source and Background Selection}\label{off_pulse}
Unlike previous studies that treated the off-pulse X-ray emission as the background, the superb angular resolution of \chandra\ can minimize the contamination from the surrounding PWN,  enabling us to investigate the off-pulse emission.  The source selection criterion for the CC mode observation, which is indicated by the orange shaded area in Figure~\ref{projected_counts}, is defined as a 4\arcsec\ wide box centered on \psr. The point spread function (PSF) generated by the raytrace simulation tool {\tt ChaRT}\footnote{\url{http://cxc.harvard.edu/ciao/PSFs/chart2/}} suggested that our source selection region contains more than 95\% of the source flux. We obtained a total of $\sim$82,000 counts in 0.5--7\,keV.  

In the projected count profile in Figure~\ref{projected_counts}(a), a bump is clearly seen on the northwest side of the pulsar.  This is likely related to the bright clumps in the PWN \citep{GaenslerAK2002, DeLaneyGA2006}.  To avoid overestimating the background level, we chose a 4\arcsec\ wide region with 1\arcsec\ separation from the source region in the southeast as the background (indicated as the gray shaded area in Figure~\ref{projected_counts}) in the analysis.  We also tried using the average of both sides as the background in the spectral analysis and found that the results are similar. 

\begin{figure}[t]
\includegraphics[bb=-10 -10 290 280, width=0.42\textwidth, clip]{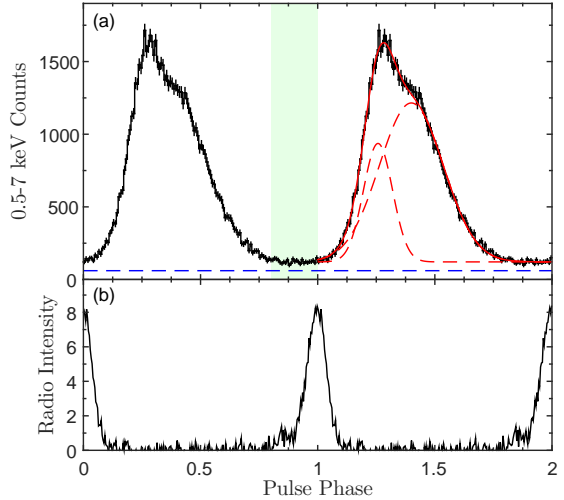} 
\caption{(a) Folded pulse profile of \psr\ in 0.5--7\,keV.  We divided the photons into 128 bins per spin period and plotted two cycles.  The green shaded region defines the off-pulse phase. The blue dashed line denotes the background level. The red solid line is the best-fit two-Gaussian model, with the two components shown by the red dashed lines. (b) The radio pulse profile in an arbitrary unit. The two profiles are aligned in phase, with phase zero corresponding to the radio peak. \label{fold_lc}}
\end{figure}

\begin{figure}[h]
\centering
\epsscale{1.1}
\plotone{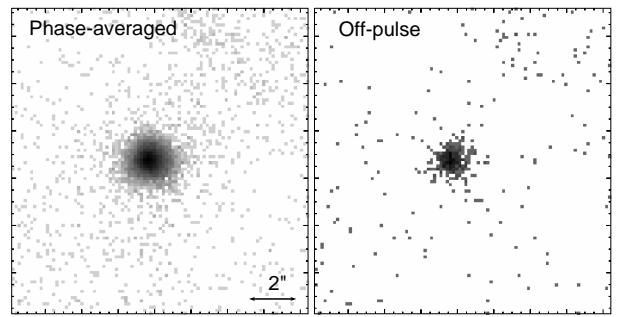} 
\caption{The HRC phase-averaged and off-pulse images of \psr\ (see Figure~\ref{fold_lc} for the definition of the phase range).  The off-pulse emission is obviously a point source consistent with the phase-averaged emission. \label{hrc_image}}
\vspace{-0.5cm}
\end{figure}

\subsection{Timing Analysis}\label{timing}
We folded the photon arrival times of the CC mode observation according to the radio timing ephemeris.  The X-ray and radio profiles are shown in Figure~\ref{fold_lc} and the radio peak is defined as phase zero. The X-ray profile shows a single, broad, and asymmetric peak, with a fast rise and a relatively slower decay, and it lags the radio peak by more than 0.2 cycles. To investigate the off-pulse emission, we collected all X-ray photons at the pulse minimum, defined as phase 0.8--1.0 (see Figure~\ref{fold_lc}), and plot the count distribution along the projected direction in Figure~\ref{projected_counts}(b). An excess of a width comparable to the model PSF is clearly seen at the pulsar position, indicating that the pulsar is still much brighter than the surrounding during the off-pulse phase.  We furthermore confirmed the detection of the off-pulse emission by examining the archival HRC off-pulse image (see Figure~\ref{hrc_image}). We searched the period of HRC data with the $H$-test algorithm \citep{deJagerRS1989} since the observation is not covered by the radio ephemeris, and then folded the photon arrival times according to the best-determined period.  The pulse profile is consistent with the CC mode profile in Figure~\ref{fold_lc}. The off-pulse HRC image is fully consistent with the pulsed image and the model PSF. 

\begin{deluxetable*}{cccccc}
\tablecaption{Best-fit Parameters for the Pulse profile of B1509 Observed with \chandra\ in Different Energy Bands. \label{profile_parameters}} 

\tablehead{\colhead{Parameter} & \colhead{0.5--7\,keV} & \colhead{0.5--1.7\,keV} & \colhead{1.7--2.7\,keV} &\colhead{2.7--4.0\,keV} &\colhead{4.0--7.0\,keV}} 
\startdata
$\mu_1$ & 0.257(1) & 0.252(3) & 0.260(3) & 0.261(3) & 0.260(3) \\
$\sigma_1$ & 0.062(2) & 0.063(5) & 0.058(3) & 0.068(4) & 0.062(2)\\
$\mu_2$ & 0.399(3) & 0.398(9) & 0.395(4) & 0.409(8) & 0.398(5)\\
$\sigma_2$ & 0.137(2) & 0.136(5) & 0.140(3) & 0.137(4) & 0.137(2) \\
F$(\rm{A}_1)$ & 0.25(1) & 0.25(4) & 0.22(2) & 0.30(3) & 0.25(2) \\
PF$_{\rm{area}}$ & 0.87(2) & 0.80(5) & 0.87(3) & 0.90(5) & 0.90(4)\\
$\chi^{2}/dof$ & 183.3/121 & 38.6/43 & 44.1/43 & 71.1/43 & 47.6/43
\enddata
\end{deluxetable*}

Previous studies found that the pulse profile can be fit with two Gaussian components \citep{KuiperHK1999, CusumanoMM2001, GeLQ2012}.  Moreover, the leading narrower peak contributes less emission at higher energy bands \citep{GeLQ2012}, and even vanishes in the 30\,MeV--1\,GeV band observed with the \fermi\ Large Area Telescope \citep{KuiperH2015}.  The two-Gaussian function is expressed as
\begin{equation}
f(\phi)=C+\frac{A_1}{\sqrt{2\pi}\sigma_1}e^{-\frac{1}{2}\left( \frac{\phi-\mu_1}{\sigma_1} \right)^2}+\frac{A_2}{\sqrt{2\pi}\sigma_2}e^{-\frac{1}{2}\left( \frac{\phi-\mu_2}{\sigma_2} \right)^2}
\end{equation}
where $f(\phi)$ is the X-ray counts at phase $\phi$, $C$ is a constant term representing the off-pulse emission, and $\mu$ and $\sigma$ are the peak location and the width of individual Gaussian functions. The normalization factor for each Gaussian function is $A/\sqrt{2\pi}\sigma$, where $A$ is the area below the individual Gaussian function.  We fit the \chandra\ 0.5--7\,keV\,profile with the above function and obtained an acceptable result with a reduced $\chi^2$ value of 1.5. The two Gaussian components peak at phase $0.257\pm0.001$ and $0.399\pm0.003$, respectively. Their phase difference is consistent with those measured with \rxte\ in 2--5\,keV \citep{GeLQ2012} and \bepposax\ MECS in  1.6--10\,keV \citep{CusumanoMM2001}.  The best-fit model and the individual components are shown in the second cycle of Figure~\ref{fold_lc} (a).  The narrower component contributes $25.1\pm1.3$\,\% of the total pulsed emission, which is in an agreement with the previous results \citep{KuiperHK1999,CusumanoMM2001}. The best-fit parameters are listed in Table~\ref{profile_parameters}. 

\begin{figure}
\epsscale{1.0}
\plotone{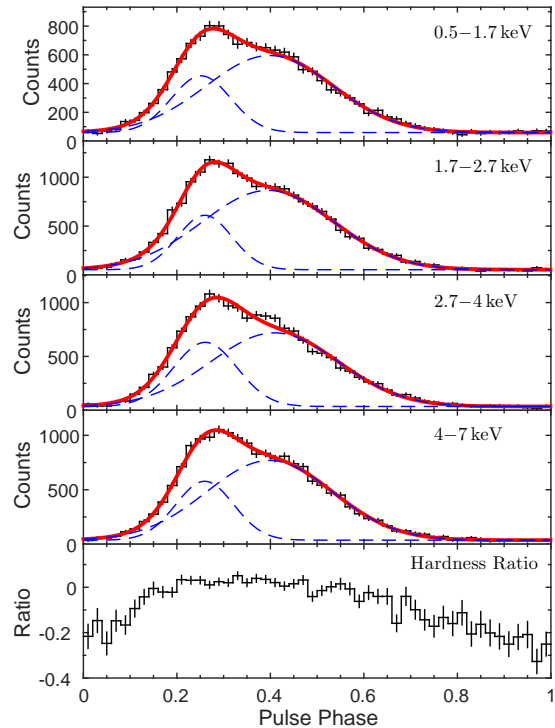} \caption{Pulse profiles of \psr\ in four energy bands (0.5--1.7, 1.7--2.7, 2.7--4, and 4--7\,keV) and the hardness ratio obtained with \chandra. The hardness ratio is defined as $(\rm{H}-\rm{S})/(\rm{H}+\rm{S})$, where $H$ is the counts in 2.7--7\,keV, and $S$ is the counts in 0.5--2.7\,keV. The best-fit model is marked by thick red lines, and the individual Gaussian components are shown by blue dashed lines.  \label{fold_lc_hardness}}
\vspace{-0.3cm}
\end{figure}

To investigate the energy dependence of the pulse profile, we divided the X-ray photons into four energy bands: 0.5--1.7, 1.7--2.7, 2.7--4, and 4--7\,keV, with a similar number of counts in each band. We then folded the X-ray light curves with 50 phase bins in each energy band, and fit the profiles with the double Gaussian function. The background-subtracted pulse profiles, the best-fit two-Gaussian functions, and the profile of each Gaussian component are shown in Figure~\ref{fold_lc_hardness}.  The best-fit parameters, including $\mu_1$, $\mu_2$, $\sigma_1$, and $\sigma_2$ are listed in Table~\ref{profile_parameters}. 

We noted that all fits have $\chi^2_\nu \sim 1.0$ except for fit at 2.7--4\,keV, which has $\chi^2_\nu \sim 1.6$. This could be contributed by the bump between the two Gaussian peaks at phase 0.4--0.45. We further divided the bands into more subbands and performed the same analysis, and found that the profile occasionally cannot be fit well. In these cases, the second peak jitters slightly and results in the bump structure around the best-fit second peak. Therefore, we concluded that the local structure is likely a statistical fluctuation. The parameters $\mu$ and $\sigma$ for each component do not significantly vary with respect to energy and are consistent with the results obtained in the similar energy bands of previous studies \citep{CusumanoMM2001, GeLQ2012}. The flux ratio $F(A_1)=A_1/(A_1+A_2)$ varies between $\sim$0.2 and $\sim$0.3 and also shows no significant energy dependency either.  This means that the pulse profile does not change evidently throughout the \chandra\ energy range, although the strength of the narrower peak decreases as the energy increases beyond the \chandra\ band \citep{KuiperH2015}.

\begin{figure}
\epsscale{1.25}
\plotone{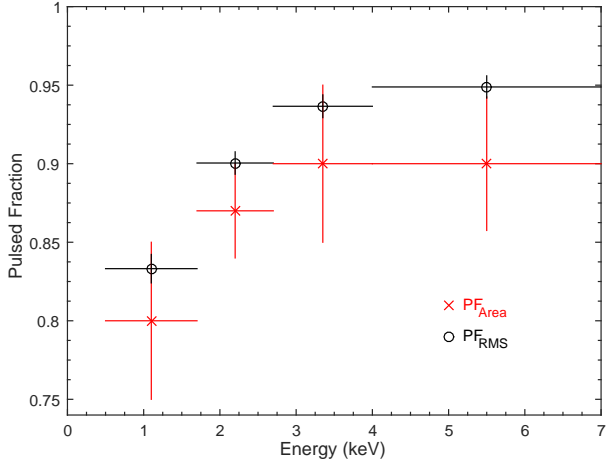} \caption{Pulsed fraction of \psr\ observed with \chandra\ ACIS estimated by the RMS method (black circles) and the area method (red crosses) \label{pulsed_fraction}}
\end{figure}

Another important derived parameter is the pulsed fraction. This can be estimated either from the pulsed and total area below the pulse profile (PF$_{\rm{area}}$) or by the root mean square (rms) method based on Fourier decomposition \citep{DibKG2009,ChenAK2015}.  The variations in PF obtained with both methods are shown in Figure~\ref{pulsed_fraction}.  We found that PF shows an increasing trend with energy, indicating that the off-pulse emission contributes more in the soft X-ray band. This gives us a hint of the softening at the pulse minimum.  The flux ratio $F(A_1)$ and PF$_{\rm{area}}$ are listed in Table~\ref{profile_parameters}. The pulsed fraction obtained with \chandra\ HRC observation is $\sim 0.88$ \citep{ChenAK2015}, which is consistent with the mean value of these four energy bands. 

The hardness ratio (HR), which is defined as
\begin{equation}
\rm HR=\frac{H-S}{H+S}\ ,
\end{equation}
where $H$ is the hard X-ray counts and $S$ is the soft X-ray counts, can provide us crude information of the spectral behavior.  Here we defined $S$ as the number of counts in 0.5--2.7\,keV, which is the combination of the two softest bands in the analysis above, and $H$ as the number of counts in 2.7--7.0\,keV.  The phase variation in HR is shown in the bottom panel of Figure~\ref{fold_lc_hardness}.  It is clear that the ratio is stable at HR$\gtrsim 0$ between phase 0.15 and 0.65, indicating that the spectral behaviors of these two Gaussian peaks are the same. Moreover, the HR drops to negative values at the off-pulse phases, which also supports the softening indicated by the PF.  This could be caused by a change in PL index with the pulse phase, or by the presence of an additional soft component (see Section \ref{spectral_analysis} below).   

\subsection{Spectral Analysis}\label{spectral_analysis}

\subsubsection{Phase-averaged Spectrum}
We first extracted the spectrum from the source and background regions described in Section \ref{off_pulse}.  The spectral fitting was carried out with the Sherpa package.  We performed the spectral fitting over the 0.5--7\,keV range and grouped the photons to at least 50 counts per energy bin. We fit a PL with the absorption model {\tt tbnew}\footnote{\url{http://pulsar.sternwarte.uni-erlangen.de/wilms/research/tbabs/}} and set the solar abundance according to \citet{WilmsAM2000}.  The best-fit parameters are $N_{\rm{H}}=(1.43\pm0.04)\times10^{22}$\,cm$^{-2}$, $\Gamma=1.18\pm0.03$, and $\chi_{\nu}^2=0.98$, where all reported uncertainties are in the 90\,\% confidence interval.  The 0.5--7\,keV absorbed flux is $(2.0\pm0.1) \times 10^{-11}$\,erg\,cm$^{-2}$\,s$^{-1}$. The $N_{\rm{H}}$ value is significantly higher than those previously reported using \chandra\ \citep{GaenslerAK2002}, \xmm\ \citep{SchockBJ2010}, and \bepposax\ \citep{CusumanoMM2001} observations.  This could be attributed to the different choice of abundance table and absorption model. We also followed previous studies to use the abundances by \citet{AndersG1989} with the absorption model {\tt phabs}, and obtained $N_{\rm{H}}=(9.7\pm0.3)\times 10^{21}$\,cm$^{-2}$, which agrees well with the \chandra\ and \bepposax\ values \citep{CusumanoMM2001, GaenslerAK2002}, and is $15\pm4$\,\% smaller than the abundance determined by \citet{SchockBJ2010}. 

\begin{figure}
\includegraphics[bb=20 15 410 345, width=0.47\textwidth, clip]{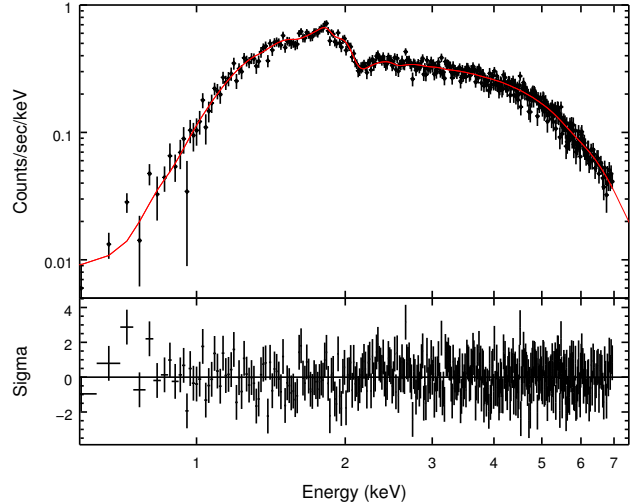} 
\caption{Pulsed spectrum of \psr. The data points in the top panel showed the 0.5--7\,keV spectrum of \psr, and the red curve corresponds to the best-fit PL model.  The bottom panel shows the fit residual. \label{on_pulse_spectrum}}
\end{figure}

\begin{deluxetable*}{cccccc}
\tablecaption{Best-fit Parameters for the PL and {\it Logpar} Models of the Pulsed X-ray Spectrum of \psr. \label{pl_logpar}} 
\tablehead{\colhead{Parameter} & \multicolumn{2}{c}{\chandra} & \multicolumn{2}{c}{\nustar} & \colhead{\bepposax} \\ 
\colhead{} & \colhead{PL} & \colhead{{\it Logpar}} & \colhead{PL} & \colhead{{\it Logpar}} & \colhead{PL} } 
\startdata
$N_{\rm{H}}$ ($10^{22}$ cm$^{-2}$) & 1.57(6) & 1.5(1) & 0.95(fixed) & 0.95(fixed) & 0.91(fixed)\\
$\Gamma$ & 1.18(4) & \nodata & 1.386(7) & \nodata & \nodata\\
$\alpha$ & \nodata & 0.9(3) & \nodata & 1.16(5) & 0.96(8) \\
$\beta$ & \nodata & 0.2(2) & \nodata & 0.11(2) & 0.16(4) \\
$\chi^{2}/dof$ & 352.7/388 & 350.4/387 & 278/254 & 254/253 & 26.6/36 
\enddata
\tablecomments{The \nustar\ and \bepposax\ results of pulsed spectra are listed for reference \citep{CusumanoMM2001, ChenAK2015}. }
\end{deluxetable*}

\begin{figure*}[ht]
\includegraphics[bb=20 15 410 330, width=0.45\textwidth, clip]{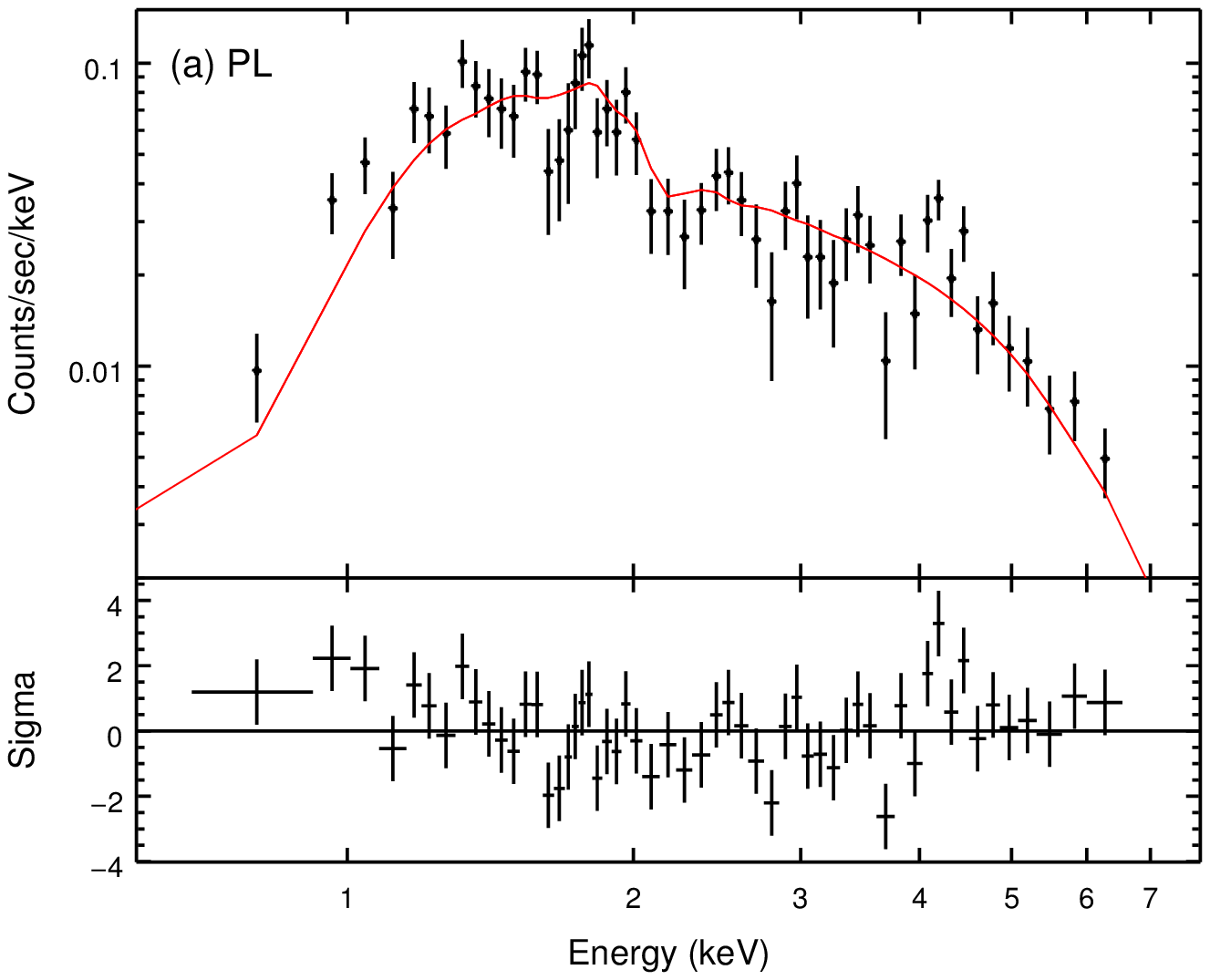}
\hspace{1.5cm}
\includegraphics[bb=20 15 410 330, width=0.45\textwidth, clip]{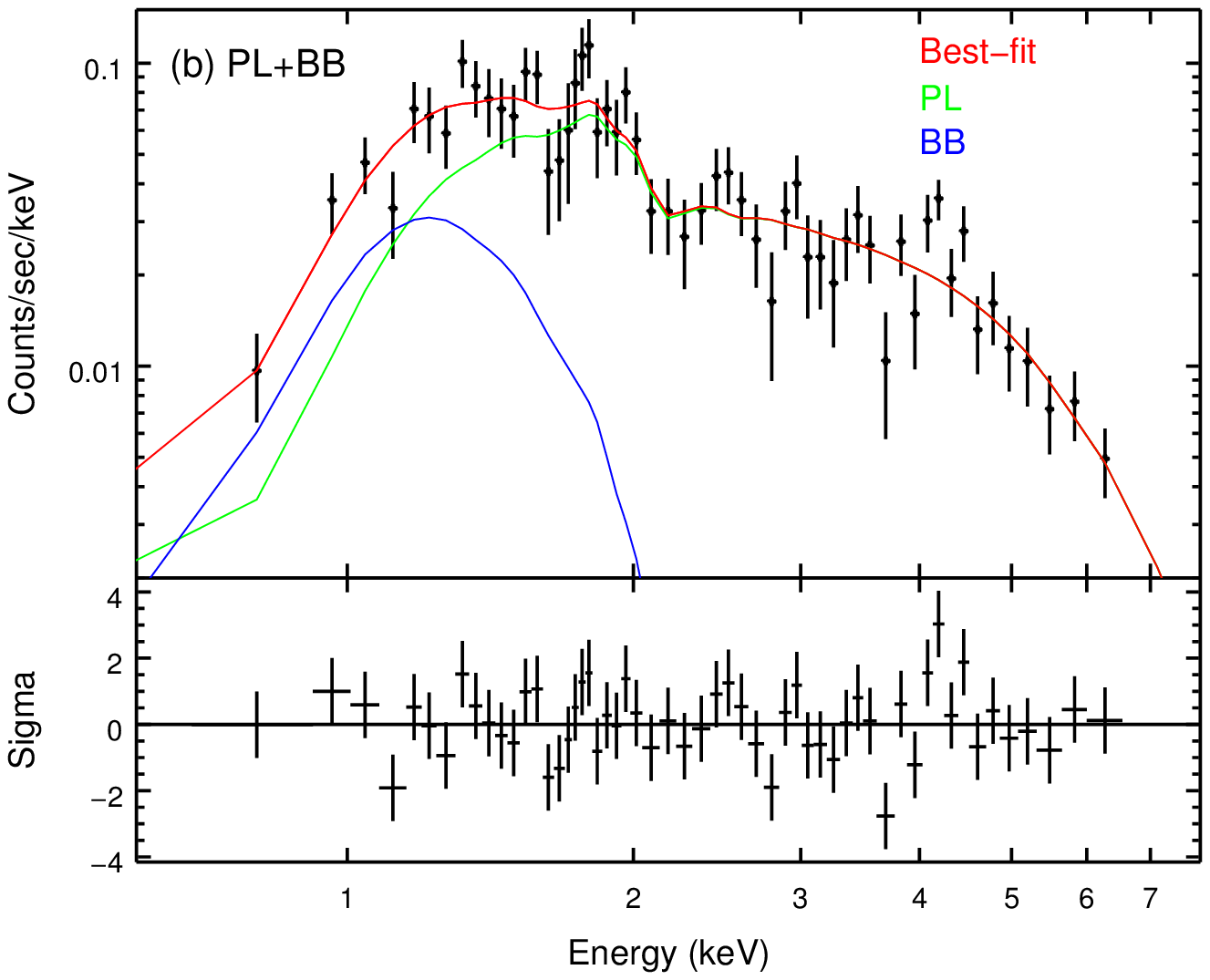} 
\caption{The \chandra\ off-pulse (phase 0.8--1.0) spectrum of \psr\ fit by (a) a PL model and (b) a PL+BB model.  \label{off_pulse_fitting}}
\end{figure*}

\subsubsection{Pulsed Spectrum}
We furthermore extracted the 0.5--7\,keV pulsed spectrum by subtracting the off-pulse emission from the pulsed one (phase 0--0.7) to compare our result with previous studies \citep{CusumanoMM2001, ChenAK2015}.  After the background subtraction, we collected $\sim$71,000 photons.  Following the same choice of abundance table and absorption model, the pulsed spectrum can be well fit with an absorbed PL, and the best-fit model with residuals is shown in Figure~\ref{on_pulse_spectrum}. The best-fit parameters are $N_{\rm{H}}=(1.57\pm0.06)\times10^{22}$\,cm$^{-2}$, $\Gamma=1.18\pm0.04$, and $\chi^2_{\nu}=0.91$. The absorbed X-ray flux within 0.5--7\,keV is $(2.5\pm0.2) \times 10^{-11}$\,erg\,cm$^{-2}$\,s$^{-1}$, where the phase duration is accounted for hereafter. The parameters are consistent with those of the phase-averaged spectrum above because the X-ray emission is dominated by the pulsed emission.

The photon index we obtained agrees with the value determined with \bepposax\ MECS \citep[$\Gamma=1.19\pm 0.04$,][]{CusumanoMM2001}, but it is significantly lower than the those \rxte\ ($\Gamma\approx1.35)$) and \nustar\ ($\Gamma=1.386\pm0.07$) values  \citep{RotsJM1998, GeLQ2012, ChenAK2015}.  Previous studies found that the photon index of \psr\ could change with energy, and the log-parabola model ({\it logpar}) gives a better fit to the broadband spectrum \citep{CusumanoMM2001, ChenAK2015}. The mathematical form of the {\it logpar} model is 
\begin{equation}
f(x)=A\left(\frac{x}{x_{\rm{ref}}} \right)^{-\alpha-\beta\log(x/x_{\rm{ref}})},
\end{equation}
where $x$ is the energy in units of keV and $x_{\rm{ref}}$ is the reference energy frozen at 1\,keV, $\alpha$ is the photon index at $x_{\rm{ref}}$, $\beta$ is the curvature term, and $A$ is the normalization. We employed this model to fit the pulsed \chandra\ spectrum, and the best-fit parameters are listed in Table~\ref{pl_logpar}. We note that the {\it logpar} model does not significantly improve the fit, and $\alpha$ is consistent with $\Gamma$ in the PL model. In addition, $\beta$ is not well determined, indicating that the spectral curvature is not obvious over the \chandra\ energy range.  We also show the \nustar\ and \bepposax\ results in Table \ref{pl_logpar} for comparison. It is clear that $\alpha$ and $\Gamma$ determined from \chandra\ are consistent with the $\alpha$ determined with \nustar\ and \bepposax.  

\begin{deluxetable}{lllcc}
\tabletypesize{\scriptsize}
\tablecaption{Best-fit Parameters for the Simultaneous Fits to the Pulsed and Off-pulse Spectra of B1509 with Different Statistics and Models. \label{off_pulse_spectral_table}}

\tablehead{\multicolumn{2}{c}{Model} & \colhead{Parameter} & \colhead{$\chi^2$ Statistic} & \colhead{Cash Statistic} \\
\colhead{Pulsed} & \colhead{Off-pulse} & \colhead{ } & \colhead{ } & \colhead{ } } 
\startdata
PL & PL & $N_{\rm{H}}$ ($10^{22}$ cm$^{-2}$) & 1.54(6) & 1.55(6) \\
 & & $\Gamma_{\textrm{pulsed}}$ & 1.16(3) & 1.16(3) \\
 & & $\Gamma_{\textrm{off-pulse}}$ & 1.85(15) & 1.78(13) \\
 & & statistic$/dof$ & 430.9/442 & 1000.7/883 \\
\tableline
PL & PL+BB & $N_{\rm{H}}$ ($10^{22}$ cm$^{-2}$) & 1.57(6) & 1.58(6) \\
 & & $\Gamma_{\textrm{pulsed}}$ & 1.18(4) & 1.18(3)  \\
 & & $\Gamma_{\textrm{off-pulse}}$ & $1.5^{+0.2}_{-0.3}$ & 1.4(2) \\
 & & $kT$ (keV) & $0.17_{-0.05}^{+0.06}$ & $0.16_{-0.04}^{+0.05}$ \\
 & & $R_{\rm{BB}}$ (km) & $9_{-5}^{+38}$ & $10_{-5}^{+39}$ \\
 & & statistic$/dof$ & 413.7/440 & 979.9/881 \\
\tableline
PL & PL+BB & $N_{\rm{H}}$  ($10^{22}$ cm$^{-2}$) & 1.57(6) & 1.58(5) \\
 & & $\Gamma_{\textrm{pulsed}}$ & 1.18(4) & 1.18(3)  \\
 & & $\Gamma_{\textrm{off-pulse}}$ & 1.6(2) & 1.5(2) \\
 & & $kT$ (keV) & $0.147_{-0.01}^{+0.007}$ & $0.148_{-0.009}^{+0.007}$ \\
 & & $R_{\rm{BB}}$ (km) & 13 (fixed) & 13 (fixed) \\
 & & statistic$/dof$ & 414.2/441 & 980.2/882 
\enddata
\tablecomments{We did not attempt to add the blackbody component to the pulsed spectrum because it is dominated by the nonthermal emission and the thermal component is subtracted out by the off-pulse emission.}
\end{deluxetable}

\subsubsection{Off-pulse Spectrum}\label{off_pulse_fit}
The off-pulse spectrum, which has never been investigated before, can be studied using our new \chandra\ observation.  Following the same source and background selection procedure, we collected $\sim$1900 net counts in the energy range of 0.5--7\,keV.  Because the  Galactic absorption should not change with pulse phase, we fit the pulsed and off-pulse spectra simultaneously with tied $N_{\rm{H}}$. We first tried a single PL model, and obtained a best-fit off-pulse $\Gamma=1.85\pm0.15$, which is obviously softer than the pulsed one ($\Gamma=1.16\pm0.03$).  Figure~\ref{off_pulse_fitting}(a) shows the off-pulse spectrum and the best-fit PL model.  The residuals show a significant curvature, especially in the low-energy end, indicating that an additional component is needed. 

\begin{figure}[t]
\includegraphics[bb=0 0 432 390, width=0.49\textwidth, clip]{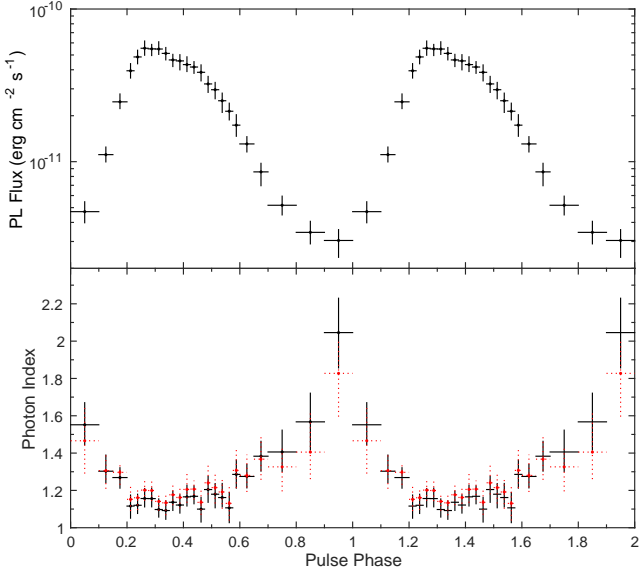} 
\caption{Phase variation of the PL flux (upper panel) and the photon index (lower panel) for \psr. The black points are obtained with a simple PL fit and the red points are obtained with PL+BB fit. All uncertainties are at 90\,\% confidence level. \label{phase_resolved}}
\end{figure}

We therefore added a blackbody (BB) component to the off-pulse spectrum and tried fitting its temperature and normalization.  We obtained $\Gamma=1.5^{+0.2}_{-0.3}$, $kT=0.17_{-0.05}^{+0.06}$\,keV, and $\chi_{\nu}^2=0.94$, which significantly improves the fit with an $F$-test null hypothesis probability of $1.3\times10^{-4}$.  Assuming a distance of 5.2 kpc, the BB radius is $R_{\rm{BB}}=9_{-5}^{+38}$\,km, not very well constrained albeit compatible with the canonical neutron star radius.  We then fixed the normalization by assuming that the emission is from the entire surface of a neutron star with a radius of 13 km.  The best-fit result yielded $\Gamma=1.6\pm0.2$, $kT=0.147^{+0.007}_{-0.01}$\,keV, and the absorbed 0.5--7\,keV fluxes of $2.2^{+0.6}_{-0.5}\times10^{-12}$\,erg\,cm$^{-2}$\,s$^{-1}$ and $8^{+20}_{-5}\times10^{-14}$\,erg\,cm$^{-2}$\,s$^{-1}$ for the PL and BB components, respectively. We note that $\Gamma$ is flatter than that obtained by fitting a single PL because part of the soft X-ray emission is contributed by the BB component.  The best-fit parameters are listed in Table~\ref{off_pulse_spectral_table}. We also used the abundances by \citet{AndersG1989} with the absorption model {\tt phabs} since the soft BB component could sensitively depend on the absorption model . The result is consistent with that obtained using {\tt tbnew}. We also attempted to use the Cash statistics \citep{Cash1979} to perform the same simultaneous fits without grouping the photons into energy bins. The result is fully consistent with that obtained from the $\chi^2$ statistics (see Table~\ref{off_pulse_spectral_table}), and the likelihood ratio test suggested that adding a BB component can significantly improve the fit at a null hypothesis probability of $6\times10^{-6}$.

\begin{deluxetable}{lcc}
\tabletypesize{\scriptsize}
\tablecaption{Variation in Photon Index of B1509 Obtained with the PL and the PL+BB Fits to the \chandra\ Phase-resolved Spectrum. \label{phase_resolved_table}} 
\tablehead{\colhead{Pulse Phase} & \colhead{$\Gamma$ (PL)} & \colhead{$\Gamma$ (PL+BB)}  } 
\startdata
0.0--0.1 & $1.55\pm0.11$ & $1.46\pm0.18$ \\
0.1--0.15 & $1.30^{+0.09}_{-0.06}$  & $1.30\pm0.10$ \\
0.15--0.2 & $1.27_{-0.05}^{+0.06}$ & $1.29_{-0.05}^{+0.08}$ \\
0.2--0.225 & $1.12\pm0.06$ & $1.15_{-0.07}^{+0.06}$ \\
0.225--0.25 & $1.12_{-0.04}^{+0.06}$ & $1.16_{-0.05}^{+0.07}$ \\
0.25--0.275 & $1.15_{-0.04}^{+0.06}$ & $1.20_{-0.05}^{+0.06}$ \\
0.275--0.3 & $1.15_{-0.04}^{+0.06}$ & $1.20_{-0.05}^{+0.06}$ \\
0.3--0.325 & $1.10_{-0.04}^{+0.06}$ & $1.14\pm0.05$ \\
0.325--0.35 & $1.09_{-0.04}^{+0.05}$ & $1.13_{-0.05}^{+0.07}$ \\
0.35--0.375 & $1.14_{-0.04}^{+0.06}$ & $1.18\pm0.06$ \\
0.375--0.4 & $1.12_{-0.04}^{+0.05}$  & $1.16_{-0.05}^{+0.06}$ \\
0.4--0.425 & $1.17_{-0.05}^{+0.06}$  & $1.20_{-0.09}^{+0.07}$ \\
0.425--0.45 & $1.17_{-0.04}^{+0.06}$ & $1.20_{-0.05}^{+0.07}$ \\
0.45--0.475 & $1.10_{-0.07}^{+0.06}$ & $1.13\pm0.07$ \\
0.475--0.5 & $1.20\pm0.07$ & $1.24_{-0.09}^{+0.06}$\\
0.5--0.525 & $1.18_{-0.07}^{+0.06}$ & $1.20_{-0.09}^{+0.08}$\\
0.525--0.55 & $1.16_{-0.05}^{+0.08}$ & $1.19_{-0.07}^{+0.09}$\\
0.55--0.575 & $1.11\pm0.08$ & $1.13_{-0.10}^{+0.09}$ \\
0.575--0.6 & $1.29_{-0.07}^{+0.08}$ & $1.30_{-0.11}^{+0.09}$ \\
0.6--0.65 & $1.27\pm0.06$ & $1.28_{-0.11}^{+0.09}$ \\
0.65--0.7 & $1.38\pm0.08$ & $1.37\pm0.11$ \\
0.7--0.8 & $1.41_{-0.10}^{+0.12}$ & $1.32_{-0.11}^{+0.13}$ \\
0.8--0.9 & $1.57\pm0.15$ & $1.41_{-0.22}^{+0.14}$ \\
0.9--1.0 & $2.05\pm0.19$ & $1.82_{-0.18}^{+0.23}$ \vspace{0.3 mm}  \\
\tableline
Pulsed & $1.18\pm0.04$ & \nodata \\
Off-pulse & $1.85\pm0.15$ & $1.5^{+0.3}_{-0.2}$ \\
Phase-averaged & $1.18\pm0.03$ & \nodata \\
\tableline
\tableline
\multicolumn{3}{c}{Tied Parameters and Statistics} \\
\tableline
Parameter & PL & PL+BB\\
\tableline
$N_{\rm{H}}$ ($10^{22}$ cm$^{-2}$) & $1.44\pm0.04$ & $1.55\pm0.05$ \\
$kT$ (keV) & \nodata & $0.142^{+0.007}_{-0.009}$\\
$\chi^2/dof$ & 2016.4/2145 & 1993.8/2144
\enddata
\tablecomments{The phase-averaged, pulsed, and off-pulse results are also listed for reference. We did not attempt to add the blackbody component to the phase-averaged spectrum because it is dominated by the pulsed emission that is well described by a power law.}
\end{deluxetable}

We also tried using a more physical neutron star atmosphere model, {\tt nsa}, to describe the soft component.  However, the fitting is not significantly better than a simple BB and the best-fit surface temperature is similar. Finally, as a consistency check, we convolved the best-fit BB+PL model with the HRC response, and this predicts 386 counts from the pulsar, which is fully consistent with the observed 390 counts. Note that the background is negligible at a level of $\lesssim 10$ photons. 

\subsubsection{Phase Variation of the Spectrum}
To investigate the phase variation of the spectrum, we divided the X-ray photons into 24 phase bins.  The bin length for the low-intensity (phase 0.7--1.1), the transition (phase 0.1--0.2 and 0.6--0.7), and the high-intensity (phase 0.2--0.6) states are 0.1, 0.05, and 0.025, respectively. We first fit all the 24 spectra simultaneously with a simple PL model and a single $N_{\rm{H}}$ value. The fit is acceptable with an $N_{\rm{H}}=(1.44\pm0.04)\times10^{22}$\,cm$^{-2}$. The variation of the PL flux and the best-fit photon index with phase is shown in Figure~\ref{phase_resolved}. $\Gamma$ stays at $\approx1.15$ between phase 0.2--0.6 and increases when the flux decreases after phase $\sim0.6$, until it reaches a maximum value of $2.0\pm0.2$ in the faintest phase bin.  We did not find significant spectral variation around phase 0.4, indicating that the bump structure in the profile (see Section \ref{timing}) is possibly a statistical fluctuation.

Similar to the analysis in Section~\ref{off_pulse_fit}, we then added a BB component across all phases and set the BB temperature and radius as free parameters.  The fit was significantly improved with an $F$-test null hypothesis probability of $4.6\times10^{-6}$, and the BB temperature is $kT=0.13\pm0.03$\,keV. However, the BB radius is $18_{-10}^{+70}$\,km, not well constrained although consistent with that in the off-pulse fit. Therefore, we fixed the BB normalization by assuming that BB emission arises from the entire surface of a neutron star. The fits yielded a consistent $N_{\rm{H}}=(1.55\pm0.05)\times10^{22}$\,cm$^{-2}$, and the BB temperature was determined as $kT=0.142^{+0.007}_{-0.009}$\,keV. After adding a BB component, the photon indices are systematically lower than those obtained from a single PL fit when the flux is low.  The photon index again shows variability, albeit slightly less significant. The results are listed in Table~\ref{phase_resolved_table}, together with the parameters for the pulsed, off-pulse, and phase-averaged spectra. 

\section{Discussion}\label{discussion}
\subsection{Thermal Emission}
\begin{figure}
\centering
\includegraphics[width=0.49\textwidth]{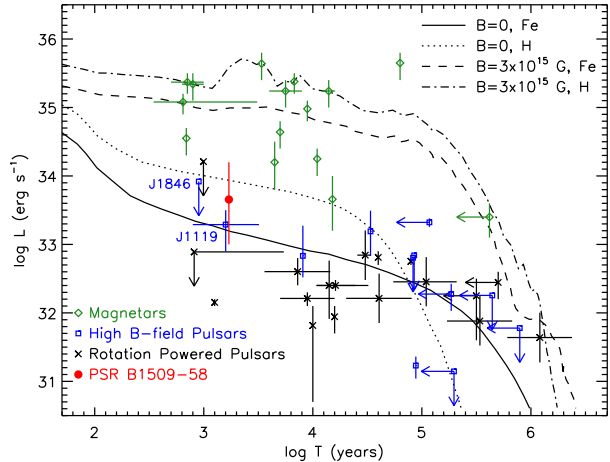} \caption{Theoretical cooling curves adopted from \citet{ViganoRP2013} for a 1.4M$_{\odot}$ neutron star with zero magnetic field and an Fe envelope (solid line), zero magnetic field and an H envelope (dotted line), strong magnetic field and an Fe envelope (dashed line), and strong magnetic field and an H envelope (dash-dotted line).  The data points show the observed thermal luminosities of magnetars (green diamond), high magnetic field RPPs (blue squares), typical RPPs (black crosses), and \psr\ (red filled circle).  The magnetar values are adopted from \citet{ViganoRP2013}, with updated ages from the McGill Magnetar Catalog \citep{OlausenK2014}. The RPP measurements and upper limits are from  \citet{ShterninYH2011}, \citet{WeisskopfTY2011}, and \citet{ViganoRP2013}. The high magnetic field RPPs are from \citet{GonzalezKL2004}, \citet{KaspiM2005}, \citet{McLaughlinRG2007}, \citet{SafiHarbK2008}, \citet{KaplanEC2009}, \citet{ZhuKG2009}, \citet{LivingstoneNK2011}, \citet{NgKH2012}, \citet{KeaneMK2013}, and \citet{OlausenZV2013}.  \label{cooling_curve}}
\end{figure}

Our spectral analysis shows evidence of thermal emission from \psr, with a BB temperature of $\sim$0.14\,keV. To compare the surface temperature with other high magnetic field RPPs, typical RPPs, and magnetars, we plotted their thermal luminosities and compared with theoretical models in Figure~\ref{cooling_curve}. Neutron stars are born as hot objects and then cool down rapidly by neutrino emission through the Urca process in the core, although different equations of state may result in a variety of cooling timescales \citep[see, e.g., ][]{NomotoT1987, LattimerPP1991, AkmalPR1998}. In general, the surface temperature of a neutron star drops to $T\lesssim 10^6$\,K after $\sim 10^3$ yr. The observational results of typical RPPs \citep[see, e.g.,][]{ShterninYH2011, WeisskopfTY2011} agree with the theoretical predictions, and they occupy the lower part of Figure~\ref{cooling_curve}.

The discovery of magnetars breaks the expectation owing to their high surface temperature of $T > 5\times 10^6$\,K and high thermal luminosities $L=10^{34}-10^{36}$\,erg\,s$^{-1}$. They are thought to be powered by the dissipation of a strong magnetic field \citep{ThompsonD1995}. A theory that unifies the magnetars and RPPs is the magneto-thermal evolution model \citep{PonsMG2009, PernaP2011, ViganoRP2013}. In this model, the decay of the crustal magnetic field causes the magnetar activities and the high surface temperature, and it is dominated by the Ohmic dissipation and the Hall drift \citep{GlampedakisJS2011}. The time evolution of the $B$-field can be described by 
\begin{equation}
\begin{split}
\frac{\partial\mathbf{B}}{\partial t} =& -\nabla\times\left\{ \frac{c^{2}}{4\pi\sigma}\nabla\times\left(\rm{e}^{\nu}\mathbf{B}\right) \right. \\
 &  \left. +\frac{c}{4\pi en_{e}}\left[\nabla\times\left(\rm{e}^{\nu}\mathbf{B}\right)\right]\times\mathbf{B}\right\}, 
\end{split}
\end{equation}
where $\mathbf{B}$ is the surface field measured by the observer at rest, $\sigma$ is the conductivity, $\rm{e}^{\nu}$ is the lapse function that accounts for the relativistic redshift, $e$ is the elementary charge, and $n_e$ is the electron number density. A stronger initial field indicates a higher energy dissipation and results in a higher surface luminosity. We adopted the theoretical cooling curves produced by \citet{ViganoRP2013}, which took all these effects into account. We plotted two cases of the initial magnetic field strengths of 0 and $3\times10^{15}$\,G. Below ages of $\sim 10^4$ yr, magnetars, high magnetic field RPPs, and typical RPPs occupy different regions in Figure~\ref{cooling_curve}. Typical RPPs are consistent with the evolutionary tracks of $B=0$ with luminosities lower than $\sim10^{33}$\,erg\,s$^{-1}$.  Magnetars, which have much higher thermal luminosities at $10^{34}$--$10^{36}$\,erg\,s$^{-1}$, are well described by the tracks of high $B$-fields. The thermal luminosity of \psr\ is comparable with other high magnetic field RPPs, e.g., PSR J1119$-$6127 with $L\approx2\times10^{33}$\,erg\,s$^{-1}$ \citep{NgKH2012}, which are in between the typical RPPs and magnetars.  This result fits the magneto-thermal evolution model well. 

Finally, we note that the return current in the magnetosphere could also heat the neutron star surface \citep{ZhangC1997, ZhangC2000}. The hard X-rays from the polar cap are reflected back to the entire stellar surface and cause a temperature of
\begin{equation}
T_s\sim 3.8\times 10^5 f_0^{1/4}P^{-5/12}B_{12}^{1/4}\rm{\,K~,}
\end{equation}
where $f_0\sim0.12$ for \psr\ is the average size of the outer gap, $P$ is the spin period, and $B_{12}$ is the surface magnetic field in a unit of $10^{12}$\,G \citep{ZhangC2000}. In the case of \psr, this effect gives a surface temperature of only $\sim0.08$\,keV, contributing a thermal luminosity lower than 9\,\% of the measured value. Hence, it is negligible in our discussion. 

In the magneto-thermal theory above, the luminosity of a neutron star is related to the magnetic field strength, which consists of the poloidal and the toroidal components. However, only the former can be inferred from the spin down, and the latter is not directly observable. Our detection of thermal emission from \psr\ completes the sample of young, high magnetic field RPPs with age $<2000$\,years. It has a similar thermal luminosity and age as other sources (namely PSRs\,J1846$-$0258 and J1119$-$6127) in the group. We therefore speculate that \psr\ has a comparable total $B$-field strength, although its dipole field is slightly weaker. High magnetic field RPPs are expected to show magnetar-like bursts as magnetars.  This idea was first suggested by \citet{KaspiM2005} and then observed in PSRs\,J1846$-$0258 and J1119$-$6127. By estimating the frequency of starquakes, \citet{PernaP2011} suggested that all types of pulsars could show magnetar-like activities, although the occurrence rate is low for lower field and aged objects. If this is true, we may expect to find magnetar-like activities in \psr\ in the sam way as in the other two high magnetic field RPPs since their ages and total $B$-fields are similar. 

\subsection{Nonthermal Emission}
Another intriguing feature of \psr\ is that the X-ray spectrum becomes softer when the flux decreases from the pulse peak.  Similar behavior has been found in other young pulsars, including the Crab \citep{WeisskopfTY2011, GeLQ2012}, PSR\,J1930+1852 \citep{LuWG2007}, and PSR\,B0540$-$69 \citep{HirayamaNE2002,GeLQ2012}. Of these, the Crab pulsar is the best-studied source. Its profile shows a double-peaked feature from radio to $\gamma$-ray bands \citep{ToorS1977, PravdoS1981}. The variation in photon index between the two peaks has been interpreted by both the outer-gap model \citep{ZhangC2002} and a phenomenological two-component model \citep{MassaroCL2000}. A recent \chandra\ observation discovered a large photon index of the Crab's off-pulse emission \citep{WeisskopfTY2011}, which cannot be explained by any previous models.

Unlike the Crab, the other pulsars show single-peaked X-ray profiles, and the profiles of both PSRs\,B1509$-$58 and B0540$-$69 can be fit with two Gaussians.  Taking this feature into account, their X-ray spectra are hardest between the two Gaussian peaks \citep{GeLQ2012}, which qualitatively agrees with the behavior of the Crab if the two Gaussians are considered as the two peaks. Moreover, these two pulsars exhibit significant softening of the off-pulse emission similar to that of the Crab pulsar \citep{HirayamaNE2002}. For PSR\,J1930+1852, this was only inferred from the HR \citep{LuWG2007} and could possibly be due to contamination by the BB emission. Therefore, further deep observation is needed to investigate its spectral behavior. 

We can qualitatively explain the phase variation of the photon index of \psr\ using the outer-gap model. In this model, the discharged pairs in the pulsar magnetosphere produce gamma-rays, which subsequently create secondary pairs that give synchrotron emission in the X-ray band. The viewing angle of \psr\ only allows us to detect the synchrotron radiation of the pairs created by the incoming gamma-rays \citep{WangTC2013}. The created pairs are quickly cooled down to the lowest Landau level via  synchrotron emission, with the photon energy given by
\begin{equation}
E_{syn}\sim 3h\gamma^2B/4\pi m_ec\sim 170{\rm MeV}(\gamma/10^3)^2(B/10^{10}{\rm G})\rm{,}
\end{equation}
with $\gamma$ being the Lorentz factor of the pairs and $B$ being the magnetic field strength at the pair-creation site. The spectrum has a high-energy cutoff of $E_{max}\sim170{\rm\,MeV}(B/10^{10}{\rm\,G})$ owing to the maximum energy of the created pair with $\gamma\sim1000$. This is much higher than the \chandra\ energy range. In addition, a low-energy turnover is present due to the cooled-down pair with a final Lorentz factor of $\gamma\sim1$ as $E_{min}\sim170{\rm\,eV}(B/10^{10}{\rm\,G})$. Below this, the spectrum is described by the low-energy tail of the synchrotron radiation, which has an asymptotic form of a PL with $\Gamma\sim 2/3$ \citep{RybickiL1986}. Between these two critical energies, the spectrum has $\Gamma\sim 1.5$ determined by the energy distribution of the pairs. Hence, the photon index in the X-ray band is controlled by the value of $E_{min}$, which is proportional to the magnetic field strength.

According to this model, the pair-creation process of \psr\ has two origins. The first origin is magnetic-photon pair creation, which occurs near the magnetic pole because a stronger magnetic field is needed, thus the synchrotron emission is observed at a narrower viewing angle. As the $B$-field is strong near the pole, $E_{min}$ is lower and hence this emission has a flatter spectrum. The pulsed X-ray emission of \psr\ is dominated by this process. The other process is photon-photon pair creation, which is distributed more isotropically in the entire magnetosphere, and $E_{min}$ is located in optical/UV bands owing to the lower magnetic field. The off-pulse emission is dominated by this process, and the X-ray emission is much softer than the pulsed one. Consequently, the variation in spectral index can be qualitatively explained by the various amounts of contribution from these emission processes across the rotational phase. 

\section{Summary}\label{summary}
We performed a detailed timing and phase-resolved spectral analysis of the high magnetic field RPP \psr\ using high-resolution \chandra\ observations. The pulse profile can be fit by two Gaussian components, and the pulsed fraction is energy dependent.  The 0.5--7\,keV pulsed spectrum is well described by a PL with a photon index of 1.18, lower than the indices at higher energy bands. This result is consistent with the broadband curved {\it logpar} spectrum. The off-pulse spectrum can be fit by a PL plus BB model. The photon index is significantly higher, and the BB has a temperature of $\sim$0.14\,keV. The thermal luminosity of \psr\ we obtained is comparable with the luminosities of two other young high magnetic field RPPs, J1846$-$0258 and J1119$-$6127, and it is brighter than the luminosities of normal RPPs. This result supports the magneto-thermal evolution model, indicating that these sources belong to the transition class of neutron stars between magnetars and RPPs and could show magnetar-like activities.  We also found that the nonthermal emission in the off-pulse phase has a much softer spectrum.  This behavior is similar to other young and energetic pulsars, including the Crab, PSRs\,B0540$-$69 and J1930+1852.  We interpreted it qualitatively using the outer-gap model, in which the pulsed emission of \psr\ is dominated by the magnetic-photon pair creation near the stellar surface, while the off-pulse nonthermal emission is dominated by the isotropic photon-photon pair creation. 

\acknowledgments
We thank the referee for the comments that improved this paper. The scientific results reported in this article are based on observations made by the {\it Chandra X-ray Observatory} and data obtained from the {\it Chandra Data Archive}. This research has made use of software provided by the Chandra X-ray Center (CXC) in the application packages CIAO, ChIPS, and Sherpa. The Parkes radio telescope is part of the Australia Telescope National Facility, which is funded by the Commonwealth of Australia for operation as a National Facility managed by Commonwealth Science and Industrial Research Organization (CSIRO). C.-P.~H. and C.-Y.~N.~are supported by a GRF grant of Hong Kong Government under HKU 17300215P. J.~T. is supported by the NSFC grants of China under 11573010.

\emph{Facilities:} CXO (ACIS, HRC), Parkes

\emph{Software:} CIAO \citep{FruscioneMA2006}, Sherpa \citep{FreemanDS2001}


\end{document}